\documentclass[12pt]{article}

\usepackage[T1]{fontenc}
\usepackage[latin9]{inputenc}
\usepackage{color}
\usepackage{babel}
\usepackage{graphicx}
\usepackage{xcolor}
\usepackage[]{hyperref}

\newcommand{\be}{\begin{equation}}
\newcommand{\ee}{\end{equation}}
\newcommand{\bc}{\begin{center}}
\newcommand{\ec}{\end{center}}
\newcommand{\ben}{\begin{eqnarray}}
\newcommand{\een}{\end{eqnarray}}

\newcommand{\la}{\langle}
\newcommand{\ra}{\rangle}

\newcommand{\kket}[1]{\vert{#1}\rangle\rangle} 
\newcommand{\bbra}[1]{\langle\langle{#1}\vert} 
\newcommand{\bbracket}[2]{\langle\langle{#1}\vert{#2}\rangle\rangle} 
 
\newcommand{\ctL}{{\cal L} }

\makeatletter
\let\myTOC\tableofcontents
\renewcommand\tableofcontents{%
  \frontmatter
  \pdfbookmark[1]{\contentsname}{}
  \myTOC
  \mainmatter }

\@ifpackageloaded{babel}{
 \AtBeginDocument{\renewcommand{\ref}[1]{\mbox{\autoref{#1}}}}
 \addto\extrasspanish{%
  \renewcommand*{\equationautorefname}[1]{}%
 }
}{}

\makeatother

\begin{document}
\title{Dissipative Quantum Hopfield Network: A numerical analysis}
\author{Joaqu{\'i}n J. Torres and Daniel Manzano\footnote{Author to whom any correspondence should be addressed (manzano@onsager.ugr.es).}\\
Institute Carlos I for Theoretical and Computational Physics\\
Electromagnetism and Matter Physics Department \\
University of Granada, E-18071 Granada, Spain.}

\maketitle
\vspace{-1cm}
\begin{abstract}
  We present extensive simulations of a quantum version of the Hopfield Neural Network to explore its emergent behavior. The system is a network of $N$ qubits oscillating at a given $\Omega$ frequency and which are coupled via Lindblad jump operators built with local fields $h_i$ depending on some given stored patterns. Our simulations show the emergence of pattern-antipattern oscillations of the overlaps with the stored patterns similar (for large $\Omega$ and small temperature) to those reported within a recent mean-field description of such a system, and which are originated deterministically by the quantum term including $s_x^i$ qubit operators. However, in simulations we observe that such oscillations are stochastic due to the interplay between noise and the inherent metastability of the pattern attractors induced by  quantum oscillations, and then are damped in finite systems when one averages over many quantum trajectories. In addition, we report the system behavior for large number of stored patterns at the lowest temperature we can reach in simulations (namely $T=0.005\, T_C$). Our study reveals  that for small-size systems the quantum term of the Hamiltonian has a negative effect on storage capacity, decreasing the overlap with the starting memory pattern for increased values of $\Omega$ and number of stored patterns. However, it also impedes the system to be trapped for long time in mixtures and spin-glass states. Interestingly, the system also presents a range of $\Omega$  values for which, although the initial pattern is destabilized due to quantum oscillations, other patterns can be retrieved and remain stable even for many stored patterns, implying a quantum-dependent nonlinear relationship between the recall process and the number of stored patterns. 
\end{abstract}

\section{Introduction}

For several decades, extensive research has been conducted in the field of artificial neural networks (ANN). These mathematical algorithms feature a networked architecture inspired by the structure of biological neural systems, aiming to emulate the way these systems process information and develop cognitive functions. Recently, there has been growing interest in extending this research into the quantum domain, driven by advancements in quantum computer development. The primary goal is to design quantum neural networks (QNN) models that can induce new intriguing emerging phenomena compared with their classical counterparts. Some of these phenomena could  be used for specific tasks, such as sequential pattern recognition or episodic memory, or be suitable for machine learning with the developing new paradigms of quantum algorithms performing better that their classical versions. This has been recently achieved for some particular cases \cite{shor:siamjc97,grover_96}. Consequently, quantum versions of classical ANNs, such as perceptrons \cite{wiebe:npj16,Benatti2019,silva:nn20,pechal:arxiv21} and Hopfield's attractor neural networks \cite{rotondo:jpa18,rebentrost:pra2018, fiorelli:njp22,marsh:prx21}, intriguing single driven dissipative quantum oscillators with associative memory capabilities \cite{labay2023}, and biologically-inspired interacting quantum neurons \cite{torres:njp22}, have been proposed in the last years. The primary focus is to determine whether these quantum systems and quantum neural networks can improve classification and pattern recognition properties compared to classical systems and neural networks or induce intriguing computational capabilities not yet described.

The Hopfield model \cite{hopfield_82} is one of the most extensively studied models in the ANN field \cite{torres:nc04,torres:fcn13,marullo:entropy20}. Its ability to store specific network activity configurations as stable attractors makes it an ideal model for theoretical studies of memory acquisition, consolidation, and recall processes. This versatility renders it suitable for various machine learning applications, such as pattern recognition. Additionally, recent model extensions that incorporate biophysical processes at the synapse or neuron levels have facilitated a theoretical understanding of dynamic or episodic memories in actual neural systems \cite{pantic:nc02,torres:nc07} and other interesting computational properties \cite{torres:fcn13}.

Due to the intriguing properties of the Hopfield model, such as Hebbian learning and the formation of autoassociative memories, it has been incorporated into many classical and modern machine learning techniques. These include Boltzmann Machines \cite{torresboltzmann}, deep learning convolutional neural networks \cite{hopfieldhybrid}, reservoir computing techniques \cite{hopfieldreservoir}, and, more recently, dense associative memories \cite{dense1,dense2,dense3,dense4}, among others. The quantum physics scientific community has taken an interest in investigating whether a quantum version of such system can induce new intriguing phenomena or offer significant advantages during information processing. A quantum extension of the Hopfield model has been recently proposed \cite{rotondo:jpa18}, wherein binary neurons are replaced by qubits and quantum stochastic jumps are introduced by Lindblad jump operators based on classical transition probabilities. These probabilities rely on the classical local fields or synaptic currents each neuron receives from its neighbours, which depend on the patterns or memories stored in the system during a previous learning process. The main finding from this study is that the addition of a quantum term to the Hamiltonian unveils a complex phase diagram featuring an oscillatory phase where the system activity transiently visits a pattern and its antipattern and other stored patterns  and their antipatterns, in addition to the standard memory and non-memory phases. This result is based on a mean-field approach, but finite-system simulations are yet to be conducted.  Recent evidence has established that for any finite system, there exists a specific time threshold beyond which the accuracy of mean field equations diminishes significantly \cite{fiorelli:njp23}. This finding underscores the critical importance of conducting finite-size simulations to reliably assess the long-time behavior of such systems.

Motivated by this fact, we have performed extensive computer simulations of the quantum Hopfield neural network to investigate its emergent behavior. Our study reveals that, for small systems, individual quantum trajectories exhibit oscillations similar to those reported in the mean-field approach \cite{rotondo:jpa18} (particularly in the limit of large quantum effects and low level of stochasticity), but contrary to this work in our study oscillations are stochastic and  are damped after averaging over multiple trajectories, with short-time oscillations dampening and long-time oscillations disappearing. This finding is consistent with numerical simulations of the Lindblad equations, which show a damping of the oscillatory behavior over time and the existence of a non-oscillatory steady state for the overlap functions. Moreover, a spectral analysis of the Liouvillian in the corresponding quantum master equation for finite size networks does  not show the emergence of steady-state oscillations but an extrapolation for infinitely large systems is compatible with the existence of such oscillations as reported in \cite{rotondo:jpa18}.  We support this conclusion by analysing the gap between the steady-state eigenvalue and the first oscillating one. We conclude that this gap may decay to zero only in the limit of infinite systems.

Through numerical simulation, we have also investigated the storage capacity of the this quantum Hopfield network. Our findings demonstrate that the storage capacity of the quantum network decreases as quantum effects become more important. This decrease is measured by the capacity of the quantum network to retrieve the information encoded in one of the stored patterns when the number of such patterns is increasing. Our observations indicate that when the system's initial state is a particular pattern, the ability of the system to retrieve the information encoded in that state is lost due to the generation of quantum oscillations, and the subsequent damping of those oscillations results in the non-recall of the initial pattern. However, although the starting pattern may not be recalled,  the information encoded in other stored patterns can still be retrieved with only a few errors, indicating that still the system has the ability to retrieve some information stored in the system. This occurs in a nonlinear manner, depending on the relevant quantum parameter.

The paper is organised as follow. In  \ref{sec1}, Model and Methods, 
the quantum Hopfield neural network  is introduced and there is an explanation about  how the simulations and analysis of the present study have been carried out. The following section describes the main results reported in our work and relates our findings to previous results. Finally, the main conclusions of this work are presented in the last section. 

This study sheds light on the emergent behavior of a Hopfield-like quantum neural network when exposed to noise and increasing number of stored patterns, presenting an exciting opportunity to explore extensions of the current quantum network systematically. This involves studying other quantum effects, including the addition of Hamiltonians in a straightforward way, and examining the implications of introducing non-linear effects affecting qubits dynamics and interactions mimicking, for example, dynamical processes observed in actual neural systems like synaptic plasticity and neuronal adaptation. Such research may pave the way for developing new paradigms of quantum neural networks with intriguing computational capabilities.

\section{Model and Methods}
\label{sec1}

The model proposed in Ref. \cite{rotondo:jpa18} is constituted by a fully connected network of $N$ qubits. To simulate the dynamics of this quantum version of the Hopfield model we use a master equation in the Lindblad form that determines the time-evolution of  the density matrix of the system $\rho$ \cite{lindblad:cmp76,manzano:aip20}

\begin{equation}
\dot{\rho}=-i[H,\rho]+\sum_{i=1}^{N}\sum_{\tau=\pm}\left(L_{i\tau}\rho L_{i\tau}^{\dagger}-\frac{1}{2}\left\{ L_{i\tau}^{\dagger}L_{i\tau},\rho\right\} \right)
\equiv {\cal L} \rho,
\label{lindblad}
\end{equation}
where $H=\Omega\sum_{i}\sigma_{i}^{x}$ is the Hamiltonian that induces quantum features in the system. For simplicity, we use natural units, $\hbar=1$. The quantum Hopfield  dynamics is created by the set of jump operators

\begin{equation}
L_{i\pm}=\Gamma_{i\pm}\sigma_{i}^{\pm},\quad\Gamma_{i\pm}=\frac{e^{\pm \beta\Delta E_{i}/2}}{\left[2\cosh(\beta\Delta E_{i})\right]^{1/2}},
\label{eq:jump}
\end{equation}
where $\beta=1/T$ is the inverse of the temperature of the bath (for practical purposes and the sake of simplicity, we will consider in the following to work with dimensionless units for $T$ which amounts to fix an energy scale such that one has $T_{C}=1$, $\sigma_{i}^{x,y,z}$ are the Pauli matrices for each qubit, and $\sigma_{i}^{\pm}\equiv (\sigma_{i}^{x} \pm i \sigma_{i}^{y})$. The term $\Delta E_{i}=\sum_{j\neq i}\omega_{ij}\sigma_{j}^{z}$,
 represents the change in energy after flipping the $i$th spin. This assumption is inherited from the classical version of the Hopfield model, and it is used to reproduce the classical behavior in absence of the quantum Hamiltonian $H$. Here, 
 $\omega_{ij}$ represents the classical synaptic interactions among neurons that encode the information of the $P$ patterns of binary strings of length $N$, denoted as $\left\{ \mathbf{\xi}_i^{\mu} \right\}$ and that are defined according to the Hebb's prescription
\begin{equation}
\omega_{ij}=\frac{1}{N}\sum_{\mu=1}^{P}\xi_{i}^{\mu}\xi_{j}^{\mu}.
\end{equation}
This assumes that once a pattern is learned by the network its information content is stored in these synaptic weights in such a way that if connected neurons are in the same firing state in the pattern the corresponding synaptic weight increases and decreases if are in different firing states.

Together with the simulation of the master equation we have also used the quantum jump trajectories method \cite{molmer:josab93,plenio:rmp98}. For simulations of both the Lindblad master equation of the system and  quantum jump trajectories we have used the Quantum Toolbox in Python (Qutip) \cite{johansson:cpc13}. 

To analyse the emergent behavior of the system we evaluate the overlap functions of the system activity with the stored patterns using the overlap functions defined as
\begin{equation}
\la m_{x,y,z}^{\mu} \ra=\left<\frac{1}{N}\sum_{i}\xi_{i}^{\mu}\sigma_{i}^{x,y,z}\right>,
\end{equation}
which define the $X,Y$and $Z-$overlaps with pattern $\mu$ respectively.  We use also the absolute overlap, defined as 

\begin{equation}
\la \left| m_{x,y,z}^{\mu} \right| \ra=\left<\frac{1}{N} \left| \sum_{i}\xi_{i}^{\mu}\sigma_{i}^{x,y,z} \right|\right>,
\end{equation}
where we have used the definition of the absolute value of an operator as $\left| A \right|=\sqrt{A^{\dagger}A}$.

\section{Results}

\subsection{Classical limit}

\begin{figure}[h!]
\begin{centering}\includegraphics[width=10cm]{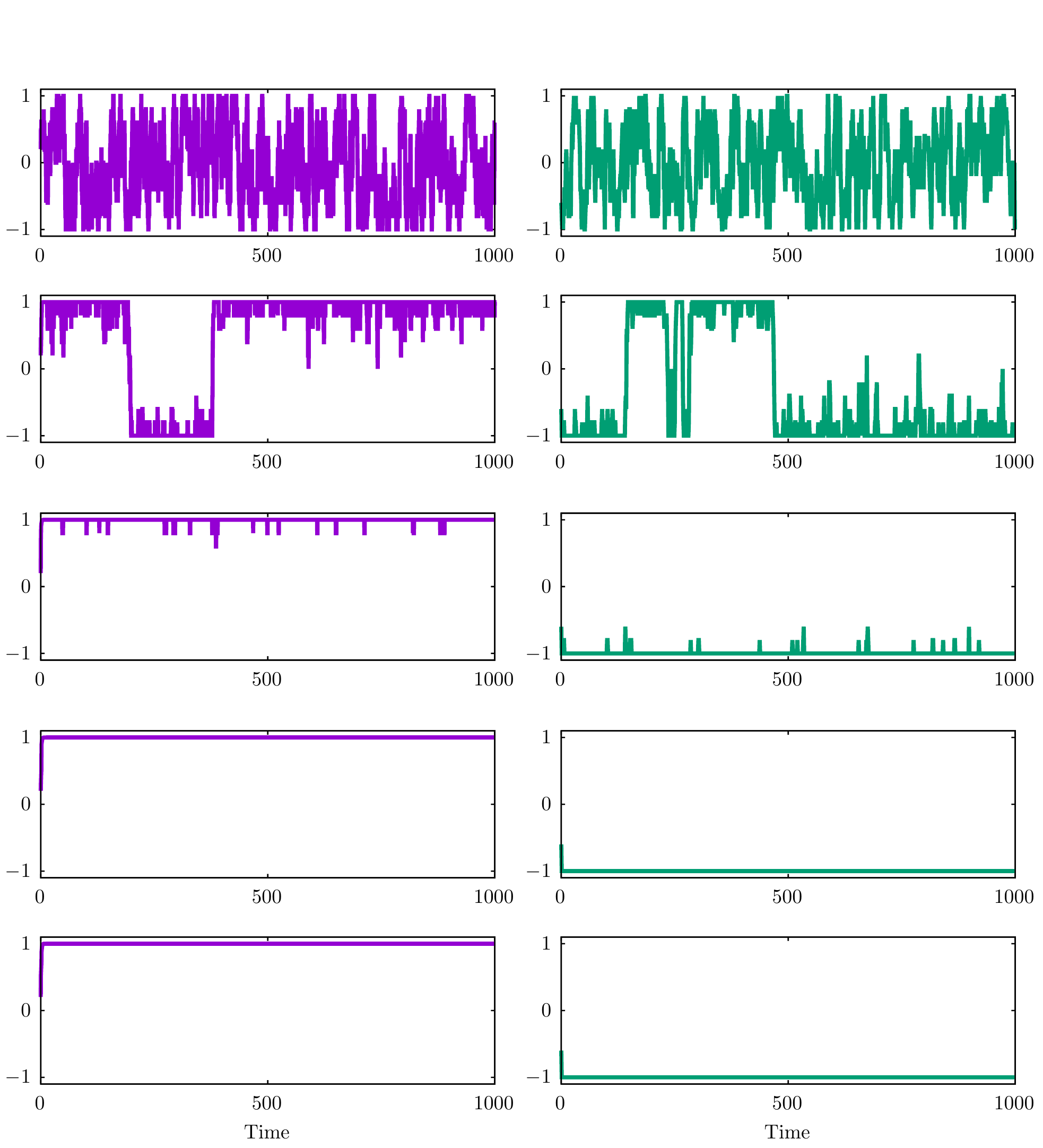}
\par\end{centering} \caption{Comparison of the retrieval properties, in terms of the overlap $m_{z}^{1}=\frac{1}{N}\sum_i\xi^1_i\sigma^z_i$ with the single stored pattern, of a QHNN with $\Omega=0$ (left, purple) and for a CHNN  (right, green). In both cases the network has $N=10$ neurons and $P=1$ stored patterns. Simulations have been performed using the quantum jump trajectories method for the QHNN and standard Monte Carlo simulation using a sequential Glauber's dynamics for the CHNN,  corresponding  to single trajectories simulations.
From top to bottom $T=1,\, 0.5,\, 0.3,\,0.1,\,0.05$. The quantum jump
method qualitatively and quantitatively reproduces the behaviour of a classical
Hopfield neural network for $\Omega\rightarrow0$. Note that the system 
can also evolve to a value $m_{z}^{1}=-1$, meaning that it has reached the stored antipattern. Simulations has been performed for time duration of $4000$ MCS.}
\label{fig1}
\end{figure}

To begin with, we have verified using different approaches that the Quantum Hopfield Neural Network (QHNN) behaves like a Classical Hopfield Neural Network (CHNN) in the limit of $\Omega\rightarrow0.$ In this scenario, the system of $N$ qubits is solely influenced by jump operators that are akin to transition or spin-flip probabilities in the CHNN.  \ref{fig1} depicts the resulting comparison between the two in single trajectory evolution of  both systems. The left panel (purple) illustrates the time behaviour of a single trajectory $Z-$overlap ($ m_{z}^{1}=\frac{1}{N}\sum_i\xi^1_i\sigma_i^z$) with one stored pattern ($P=1$)  using the quantum jump trajectories method. Meanwhile, the right panel (green) also showcases the variation of $ m_{z}^{1}$ in time for a CHNN Monte Carlo simulation using the Glauber's dynamics \cite{peretto_92}. In both cases, each panel from top to bottom were calculated with decreasing bath temperature $T$.  This single trajectory analysis gives overall information about the behaviour of the system in the presence of thermal fluctuations, the typical size of such fluctuations, the presence of attractors in the system, and the size of the energy barriers separating these attractors. Moreover, if the system is ergodic, as it is the case here for $P=1,$ a time average of order parameters (such as the $Z-$overlap) over these single trajectories will represent meaningful quantities of the system.

We want to clarify that the equivalence between  classical and quantum dynamics is expected in the limit of $\Omega\rightarrow0$ and a diagonal initial density matrix \cite{garrahan:pa18}. On the other hand, the simulation methods are very different and in this section we compare then confirming that the quantum and classical systems exhibit similar qualitative and quantitative behaviour over the entire temperature range for $\Omega=0.$

\begin{figure}[h!]
\begin{centering}
\includegraphics[width=10cm]{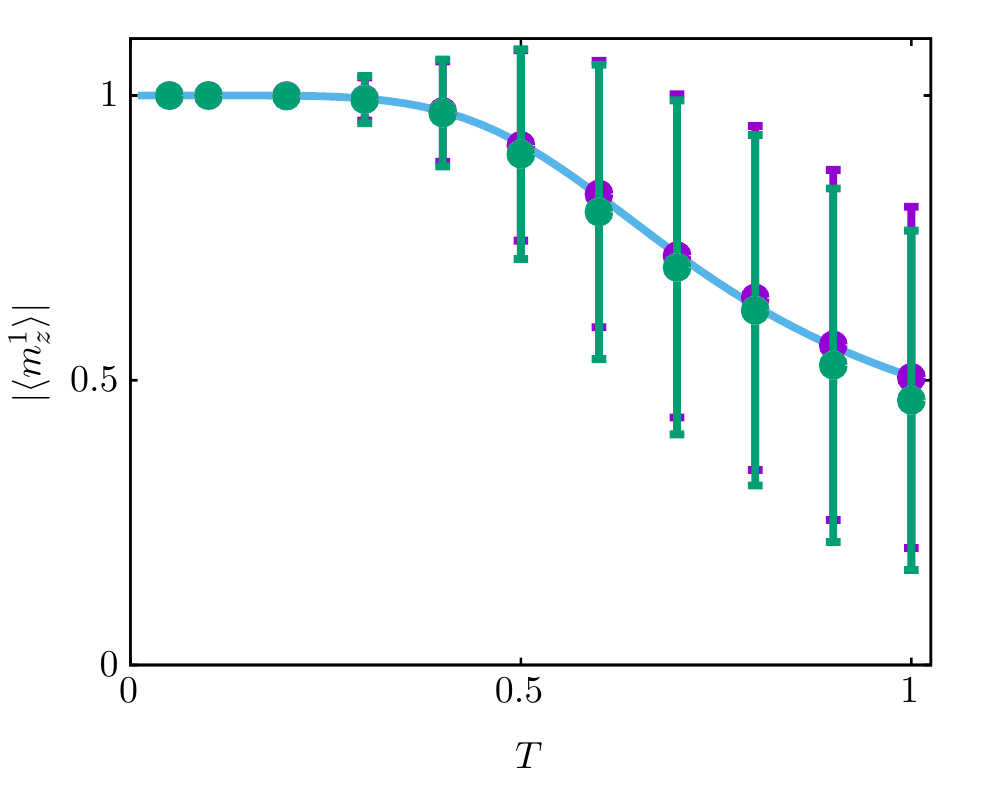}
\par\end{centering}
\caption {Absolute values of the steady state overlap with a single stored pattern ($N=10$) $\xi^1=\{1,1,1,1,1,-1,-1,-1,-1,-1\},$ as a function of the temperature, for the QHNN (blue line by integrating the master equation and purple filled circles by time averaging in a steady-state single quantum trajectory) and the CHNN (green filled circles corresponding to time averaging in the steady-state of a single Monte Carlo simulation time series using Glauber's dynamics). The error bars have been computed using the standard deviations of the absolute values of the data corresponding to the overlap time series (with a duration of $4000$ MonteCarlo steps) which also has been used to compute the points (these correspond to the mean value of the absolute value of the data in the time series). Simulations performed with a single stored random pattern depicts exactly the same behaviour, a fact that clearly indicates that the results illustrated in this figure does not depend on the particular realization of the stored pattern (data not shown).}
\label{fig2}
\end{figure}

We can also conduct this comparison in steady-sate conditions by plotting $\la |m_{z}^{\mu}| \ra$ as a function of $T$ as it is shown in \ref{fig2}. In this figure, it is displayed the average steady state of the obtained $Z-$overlap 
using integration of  \ref{lindblad} (blue line), together with the overlap of the quantum and classical Monte Carlo simulations after they reach the steady-state (purple and green dots). Note that error bars are comparable for both quantum and classical simulations.
Possible sources for slightly observed differences could be due to the different
updating scheme for quantum and classical cases and the very small
system size that is used in simulations ($N=10$). In both cases, the system
shows stochastic jumps among the pattern and antipattern attractors
when the system is near the critical temperature (which is $T_{C}=1$
for $N\rightarrow\infty$)

\subsection{Emergence of quantum oscillations}
We have  explored the possible emergence of other intriguing behaviour in the system, the existence of quantum oscillations among stored patterns as those reported in Ref. \cite{rotondo:jpa18}. In this work, the authors demonstrated, within a mean-field approach only valid for $N\rightarrow \infty,$ that there are pattern-antipattern oscillations due to the coherent quantum dynamics. In simulations or in actual quantum computers this limit can not be reached so the existence of similar type of memory oscillations with the same features in finite-size system's simulations could be a very suitable test to validate such mean-field results. In the study reported in \cite{rotondo:jpa18}, the authors identified three main phases.  If $\Omega$ and $T$ are small enough the system is in a {\it Retrieval Phase} (RP) where it recalls the pattern or antipattern configurations and remains stable in it. For bigger values of 
$\Omega$ and $T$ the system is in a {\it Paramagnetic Phase} (PP) where there is no retrieval of the stored patterns at all. Finally, there is a phase with {\it Limit Cycles} (LC) where the system presents sustained oscillations.

In general, the presence of an oscillatory phase can be better understood by analysing the spectral properties of the Liouvillian superoperator $\cal L$. If we transform the density matrix of the system 
into a vector we can write \ref{lindblad} as \cite{manzano:aip20}

\begin{equation}
\frac{d \kket{\rho(t)}}{dt} = {{\cal L }} \kket{\rho(t)},
\end{equation}
where $\kket{\rho(t)}$ is a vector in the Fock-Liouville space. In this representation, $\cal L$ is  a complex 
matrix of dimension $2^{2N}\times 2^{2N}$. As this matrix is non-Hermitian it would have both right, $\kket{\Lambda_i^R}$, 
and left, $\bbra{\Lambda_i^L}$, eigenvectors corresponding to the eigenvalues $\Lambda_i$ fulfilling 

\begin{eqnarray}
\ctL \kket{\Lambda_i^R} &=& \Lambda_i \kket{\Lambda_i^R}, \nonumber\\
\bbra{\Lambda_i^L} \ctL  &=& \Lambda_i \bbra{\Lambda_i^L}.
\end{eqnarray}
Therefore, as the set of left and right eigenvectors form a biorthogonal basis of the Hilbert space of the density matrices, the time-evolution of the state can be computed as 

\begin{equation}
\kket{\rho(t)}=\sum_i c_i e^{\Lambda_i t}  \kket{\Lambda_i^R},
\label{eq:time_evol}
\end{equation}
being $c_i$ the overlap between the initial state and the left eigenvalues $c_i=\bbracket{\Lambda_i^L}{\rho(0)}$. Evan's Theorem \cite{evans:jfa79} ensures that there is at least one eigenvalue with zero real value.  This corresponds to the steady-state of the system. It can also be demonstrated that all non-real eigenvalues appear in complex conjugated pairs. Additionally, there is still the possibility of having couples of eigenvalues with zero real part but imaginary finite values. These eigenvalues correspond to oscillating coherences, meaning states that evolve permanently \cite{albert:pra14,manzano:av18}. The presence of these states may justify the LC phase reported in Ref. \cite{rotondo:jpa18} for infinite systems. 

\begin{figure}[ht!]
\hspace*{-0,5cm} \includegraphics[width=14.3cm]{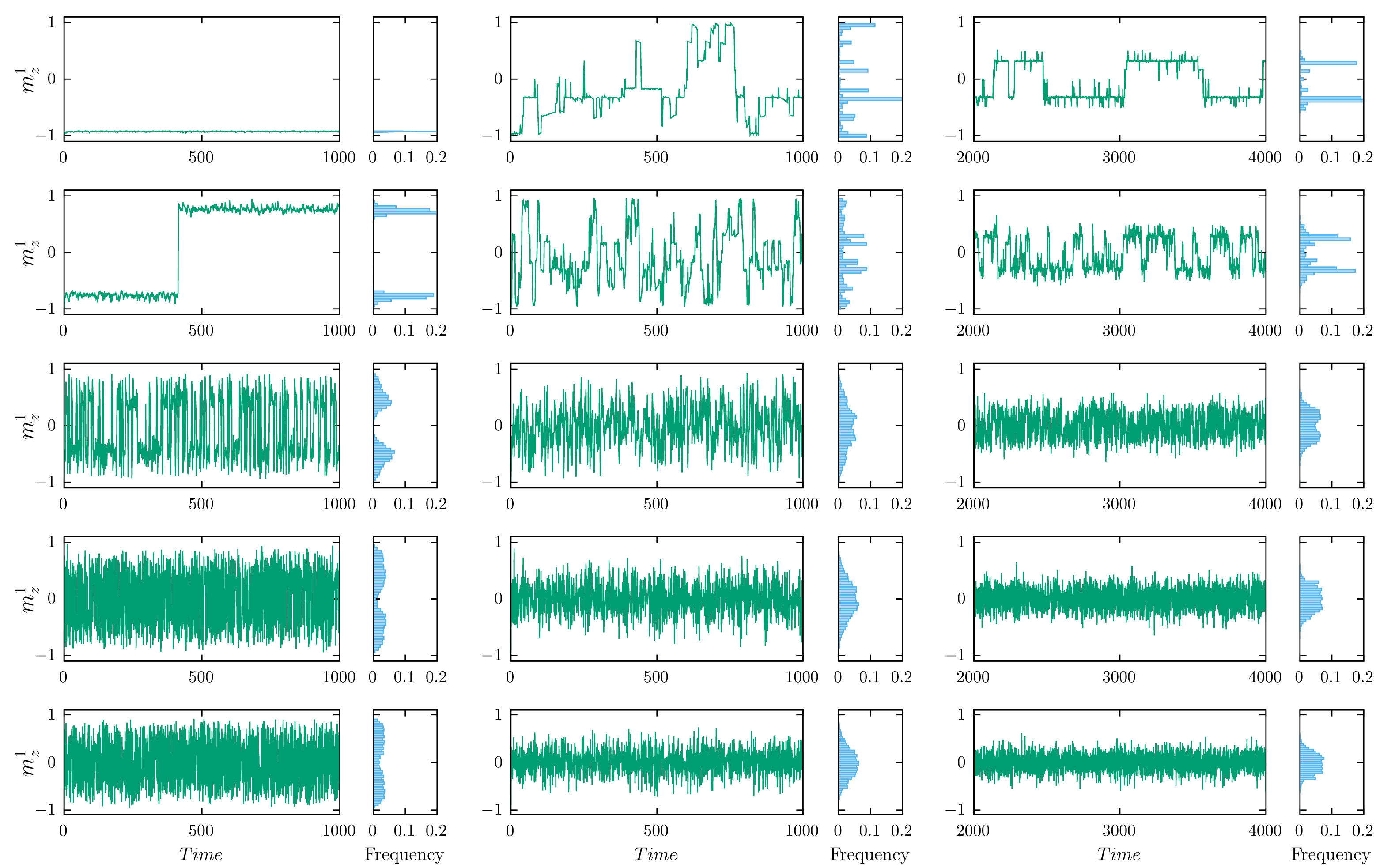}
\caption{Retrieval properties of a QHNN with $N=12$ qubits as a function of $\Omega$ for $P=1$ (left panels), $5$ (center panels), and $10$ (right panels)
stored patterns and $T=0.005$. The values of the parameter $\Omega$ are (from top to bottom in each column) $0.1,\; 0.2,\; 0.5,\; 1,$ and $2$. Green time series corresponds to the time-dependent behaviour of the single trajectory overlap functions  $m_z^\mu$ and are computed using the quantum trajectories or quantum jump method.  Note that for large $P$ the system is jumping among different metastable mixtures states together with the pattern attractors (that are also metastable). The resulting histograms (blue horizontal boxes histograms to the right of each time series) show multiple peaks histograms as it the system is in a ``non-ergodic'' phase (for $\Omega\rightarrow 0$) with multiple mixtures or spin-glass states that for this system size are metastable. The amplitude of the oscillations among the attractors decreases when $P$ increases,  making the ``non-ergodic'' like histograms more concentrated around  $m_z^\mu=0$ for larger $P.$ The system also becomes more ``ergodic''-like when $\Omega$ enlarges and quantum effects are more important which indicates the positive effect of quantum oscillations to overcome mixtures or spin-glass states. Histograms represent the probability distribution for a trajectory with $t\in[0,4000]$. }
\label{fig3}
\end{figure}

We have investigated the possible emergence of this oscillatory phenomena in finite systems by first calculating single trajectory overlaps $m_z^\mu=\frac{1}{N}\sum_i\xi^\mu_i\sigma_i^z$ of the system activity with stored patterns as a function of time for different values of the parameter $\Omega$ through the quantum jumps method \cite{molmer:josab93}. The results are illustrated in  \ref{fig3}, where $m_z^\mu$ is displayed as a function of time (green lines) along with the overlap histograms computed from the corresponding time series (blue horizontal histograms to the right of each time series). This was performed for a system of $N=12$ qubits with $P=1$ (upper left), $5$ (upper right), and $10$ (lower) stored patterns, $T=0.005$, and varying $\Omega$ values that cover both the FM and LC phases. The results demonstrate that when the quantum parameter $\Omega$ increases above zero it emerges a time-dependent state characterized by stochastic changes of the system activity among pattern and anti-pattern attractors, as well as between different patterns and mixture of patterns. When there is only one stored pattern (upper left), the histograms of the corresponding overlap time series exhibit only two peaks corresponding to the stochastic wandering of the system  around the pattern and anti-pattern configurations.  When multiple stored patterns are considered, there are not only jumps between different patterns and their corresponding anti-patterns, but also among different mixture states or other metastable states. This leads to multiple possible attractors and histograms appear with several peaks resembling the onset of a non-ergodic spin-glass phase as in the CHNN, particularly for $\Omega\rightarrow 0$. In all studied cases, when $\Omega$ increases, as quantum oscillations are more likely to occur, the corresponding histograms of the overlap time series become smoother and more concentrated around $m_z^\mu=0$, and the system becomes more ``ergodic'' like. Note that at the trajectories level there are jumps between patterns even when the system is in the retrieval phase (FM) indicating that such attractors are now metastable.

\begin{figure}[ht!]
  \begin{centering}
\includegraphics[width=10cm]{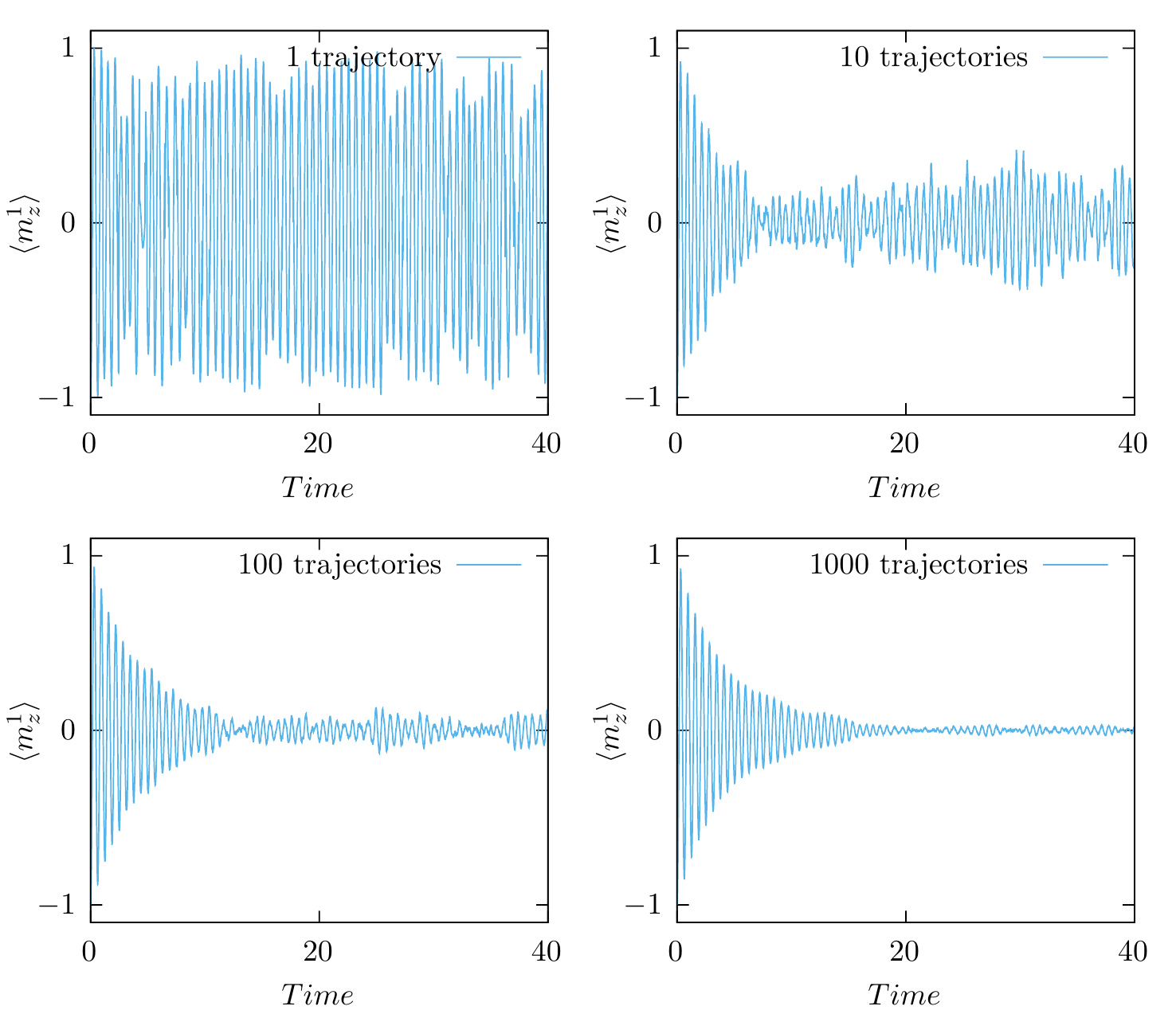}
\par\end{centering}
\caption{The effect of averaging over many trajectories in the oscillatory phase. The figure depicts that the emerging pattern-antipattern oscillations in a $N=8$ size dissipative QHNN, storing a single unbiased stored pattern ($P=1$), become damped when the number of trajectories over which the system state is averaged increases. Simulations has been done for $\Omega=5$ and $T=0.2.$}
\label{trajectoriesdamped}
  \end{figure}

\begin{figure}[ht!]
\begin{centering}
\includegraphics[width=12cm]{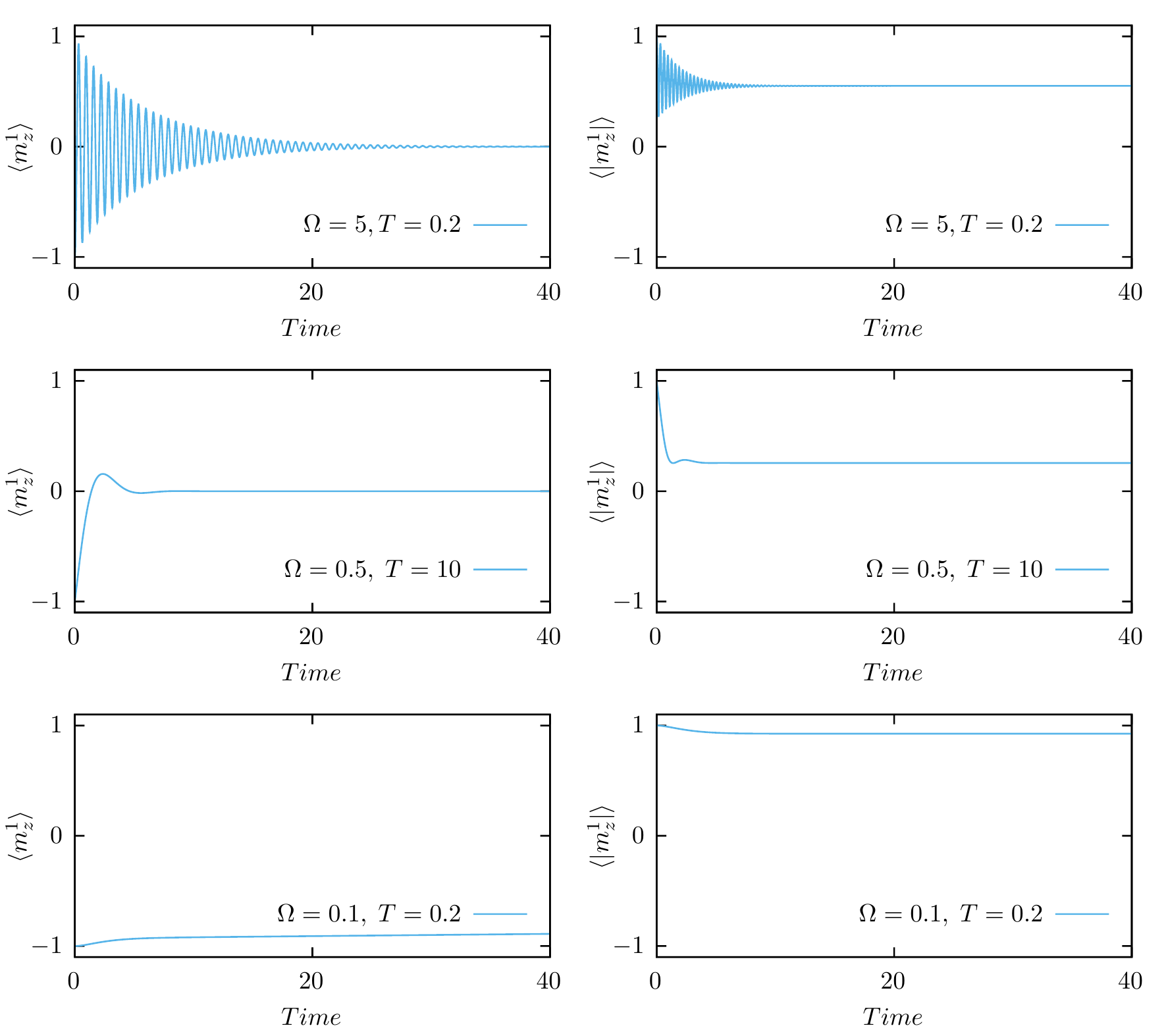}
\par\end{centering}
\caption{Time-dependent behaviour of the averaged overlap $\la m_z\ra $ (left), as well as $ \la \left| m_z^1 \right| \ra$ (right), for a system of $N=10$ qubits computed by direct integration of  \ref{lindblad}. The parameters $\Omega$ and $T$ are selected so the dynamics belong to the LC (top), PM (middle), and FM phases (bottom) in the mean-field phase diagram reported in \cite{rotondo:jpa18}. In all simulations we have considered a single unbiased pattern (i.e. $P=1$). }
\label{fig4-1}
\end{figure}

It is also interesting to remark that the above results are for single trajectory simulations. Interestingly, when one averages over different trajectories, e.g. in the oscillatory phase, oscillations start to be damped at large times, depicting a clear damping in time of the oscillations when the number of trajectories increases significantly,  as it is shown in \ref{trajectoriesdamped}. As we will see next (see also figure \ref{fig4-1}) this is equivalent to the behaviour observed using the master equation formalism that corresponds to an average over infinity number of trajectories. It is worth noting to say that for the set of parameters we are using in this figure (i.e. large $\Omega$ and small $T$) even in the single trajectory case (\ref{trajectoriesdamped} top left) transiently sustained smoother oscillations similar to those reported in the mean-field can emerge between consecutive stochastic jumps.

To study the average behaviour of the system in the different phases we have also calculated both the averaged overlap $\la m_z^1 \ra$ and its absolute value $\la \left| m_z^1 \right| \ra$ as a function of time for a density matrix starting in the anti-pattern of the system. In this case, we have studied a system of $N=10$ qubits with the parameters $\Omega$ and $T$ covering the whole phase space. The results are displayed in \ref{fig4-1} where we can see that only in the regime of LC we can observe meaningful damped oscillations. In this case (top panel),  the averaged overlap oscillates between $+1$ and $-1$, meaning that the system oscillates between the pattern and the anti-pattern configurations. For long times the final averaged overlap is zero as the system evolves to a mixture of pattern and antipattern configurations. This effect can be concluded from the average of the absolute value of the overlap that shows the system reaching a steady state with a finite overlap with the stored pattern. In comparison, for the  PM phase (middle panel) the system rapidly evolves to a state with zero average overlap -- also with a zero averaged absolute value overlap -- meaning that it does not reach any stored pattern. Finally, for the FM phase the system presents no oscillations and it just remains close to its initial state. From the results presented in \ref{fig4-1} is clear that the system presents oscillations in the LC phase but these oscillations are damped and disappear for long times.

\begin{figure}[ht!]
\begin{centering}
\includegraphics[width=10cm]{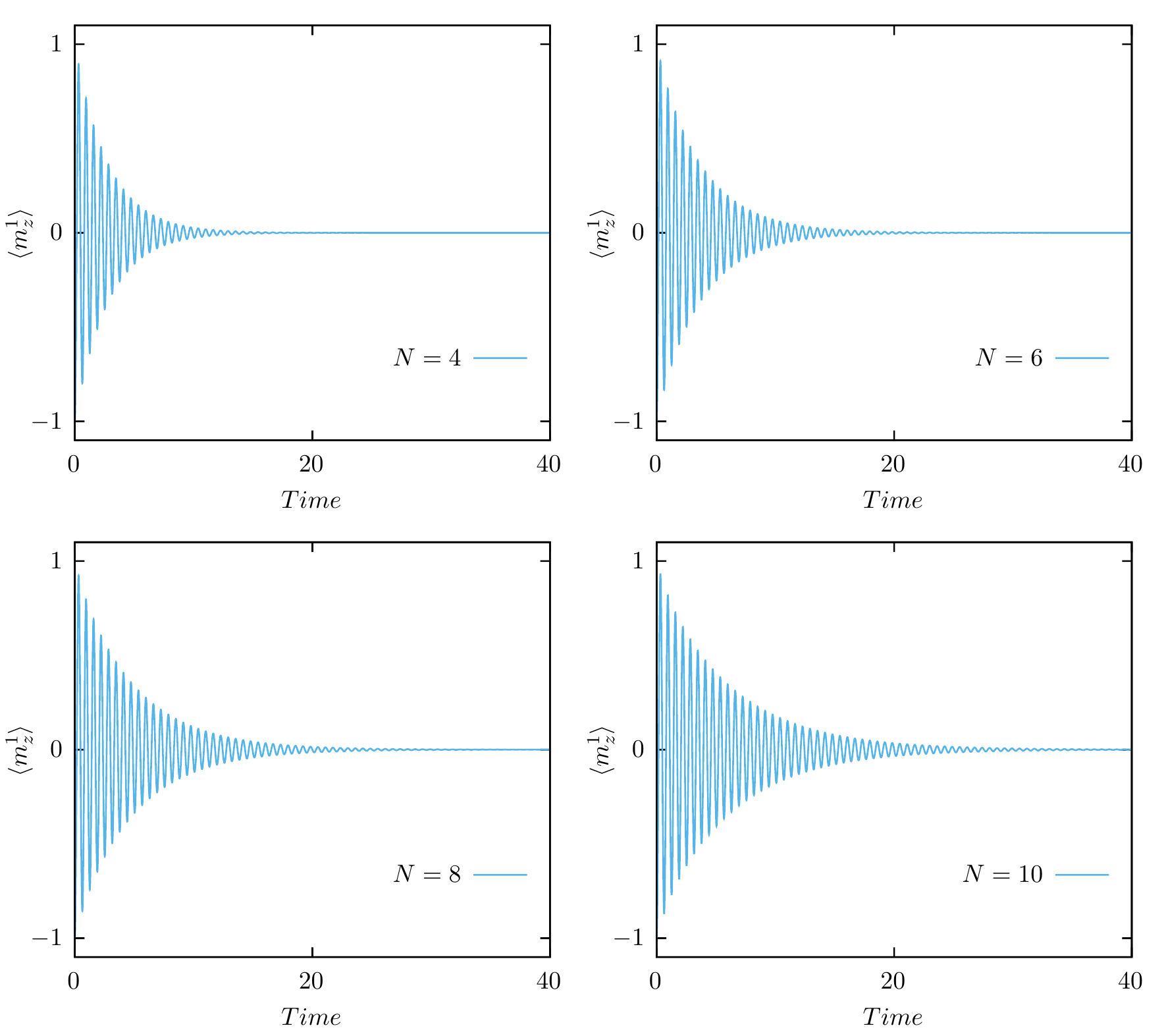}
\par\end{centering}
\caption{Oscillations for the overlap $m_z^1$ for a representative point of the LC phase ($\Omega=5$, $T=0.2$) and for different system sizes. In all simulations we have considered a single unbiased pattern (i.e., $P=1$). }
\label{fig5}
\end{figure}

\begin{figure}[ht!]
\begin{centering}
\includegraphics[width=5cm]{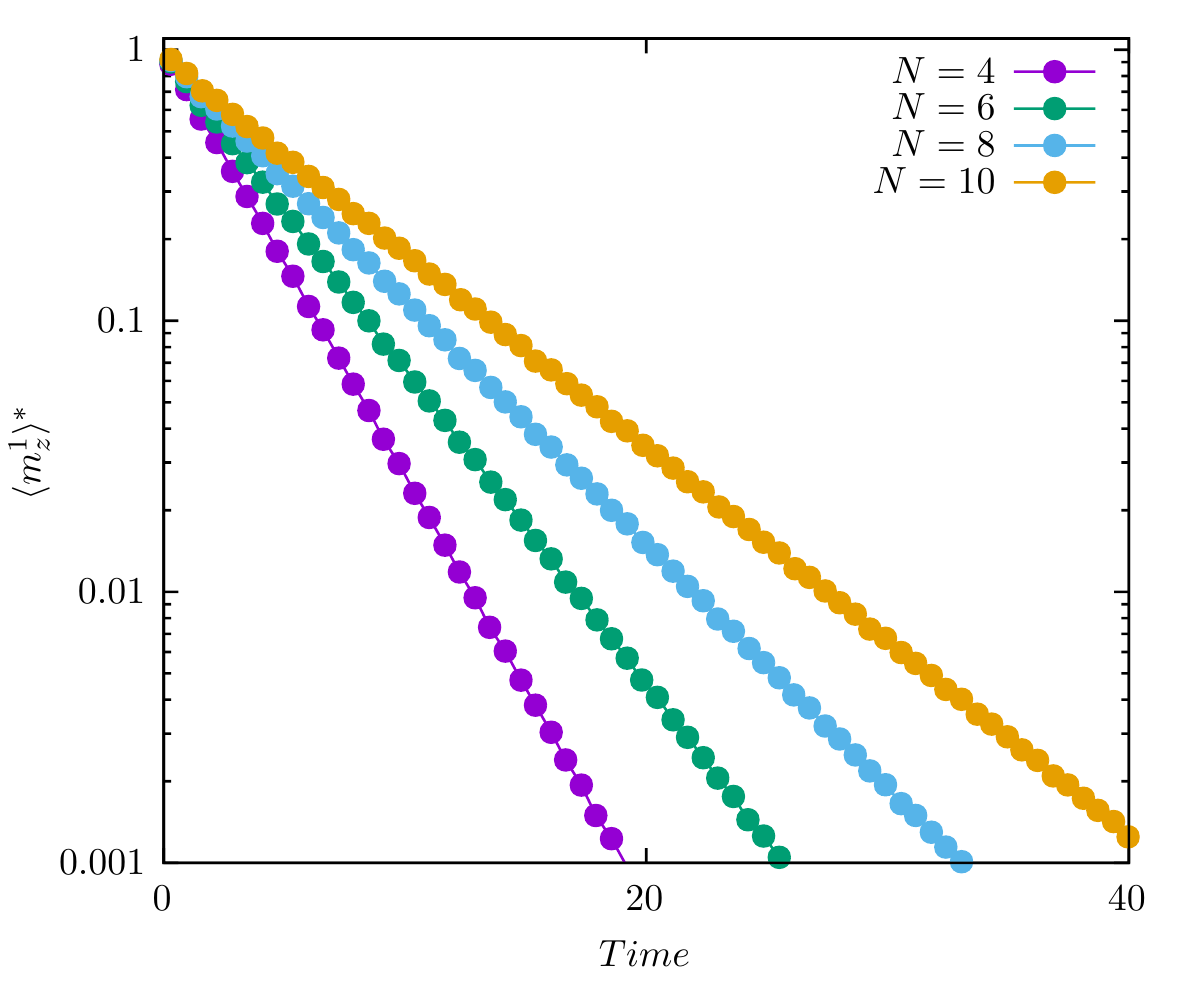}\includegraphics[width=5cm]{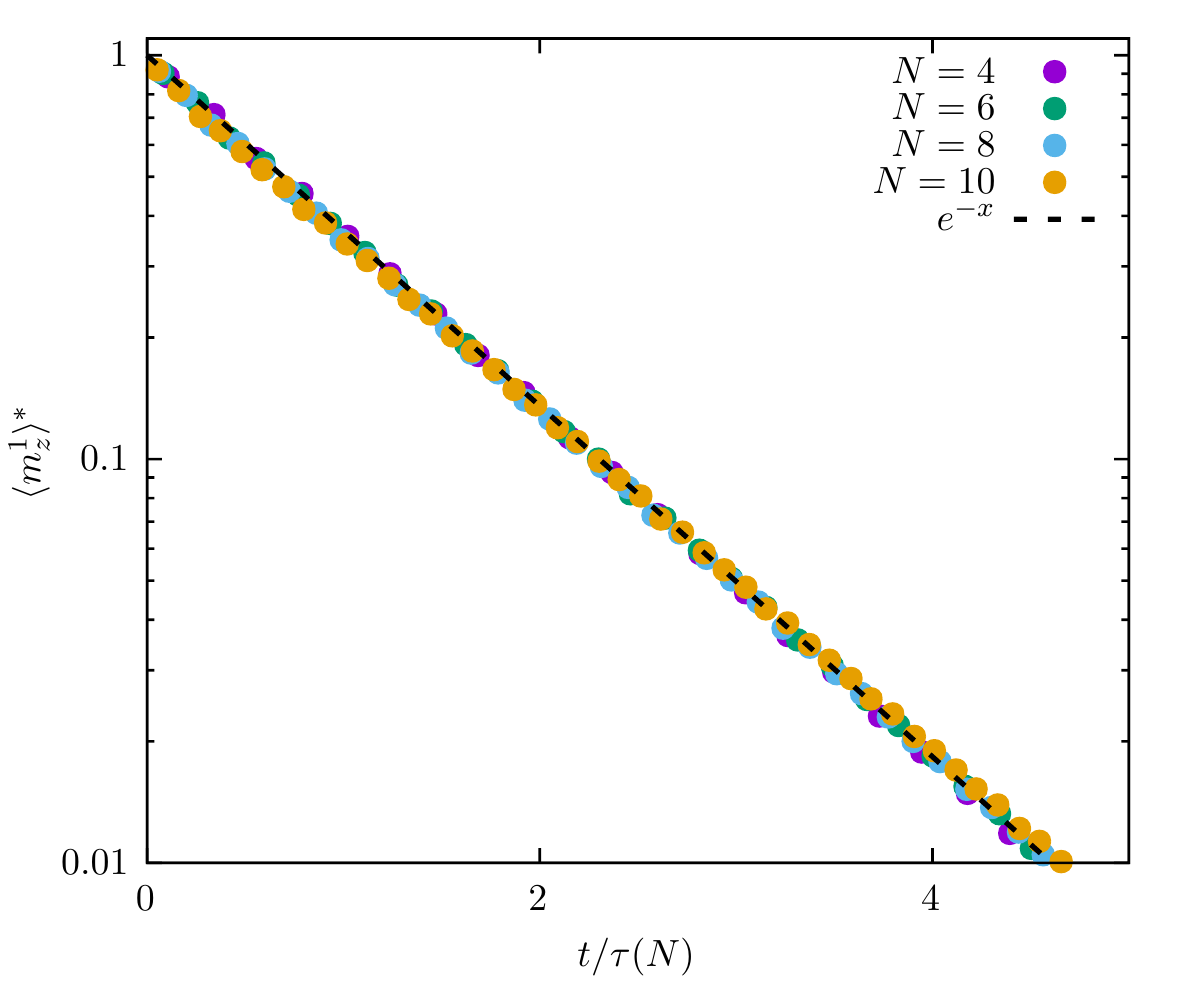}
\par\end{centering}
\caption{Left: Maximum value of the mean overlap $\la m_z^1\ra^*$ for each period of the oscillations as a function of time for systems of sizes $N=4,\, 6,\,8,$ and $10$ for the overlap curves illustrated in \ref{fig5}. Right: Collapse of the curves in the left panel showing the size scaling $\la m _z^1 \ra^*(N)=e^{-t/\tau(N)}$ with $\tau(N)=N^\delta$ with $\delta=0.75\pm 0.01.$ }
\label{fig5-2}
\end{figure}

We have also studied the oscillations of the overlap, in the LC phase, for different system sizes $N$. In \ref{fig5} we can see the overlap for different system sizes and $\Omega=5$, $T=0.2$, while in \ref{fig5-2} we have plotted the resulting relative maxima of the oscillations for each oscillation cycle, namely $\la m_z^1\ra^*$. This is done for various system size (left panel) in a semi-log scale that illustrates an exponential decay of the oscillations, with a time constant $\tau(N)$ that scales with $N$ as $\tau(N)\sim N^\delta$ with $\delta\approx 0.75$. In the right panel there is a collapse of the curves by the use of this fit. This result is consistent with the eigenvalue decomposition provided in \ref{eq:time_evol} and indicates that when the system size increases the oscillations remain for longer times, meaning that they could survive for infinite time in the limit of infinite large systems. However, for realistically sized finite systems, the oscillations will be short-lived and will not significantly impact the system dynamics.

\begin{figure}[h!]
\begin{centering}
\includegraphics[width=5cm]{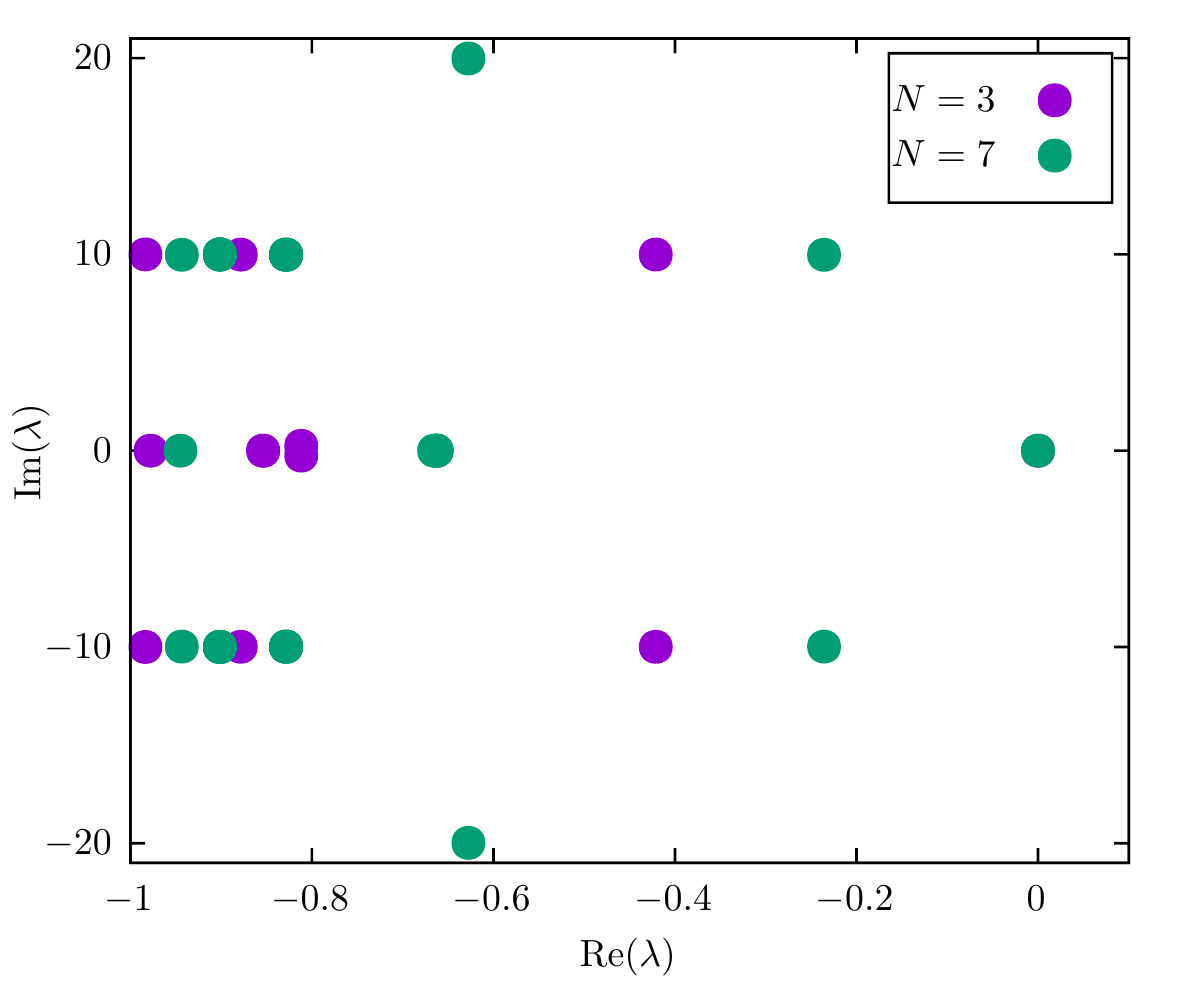}\quad \includegraphics[width=5cm]{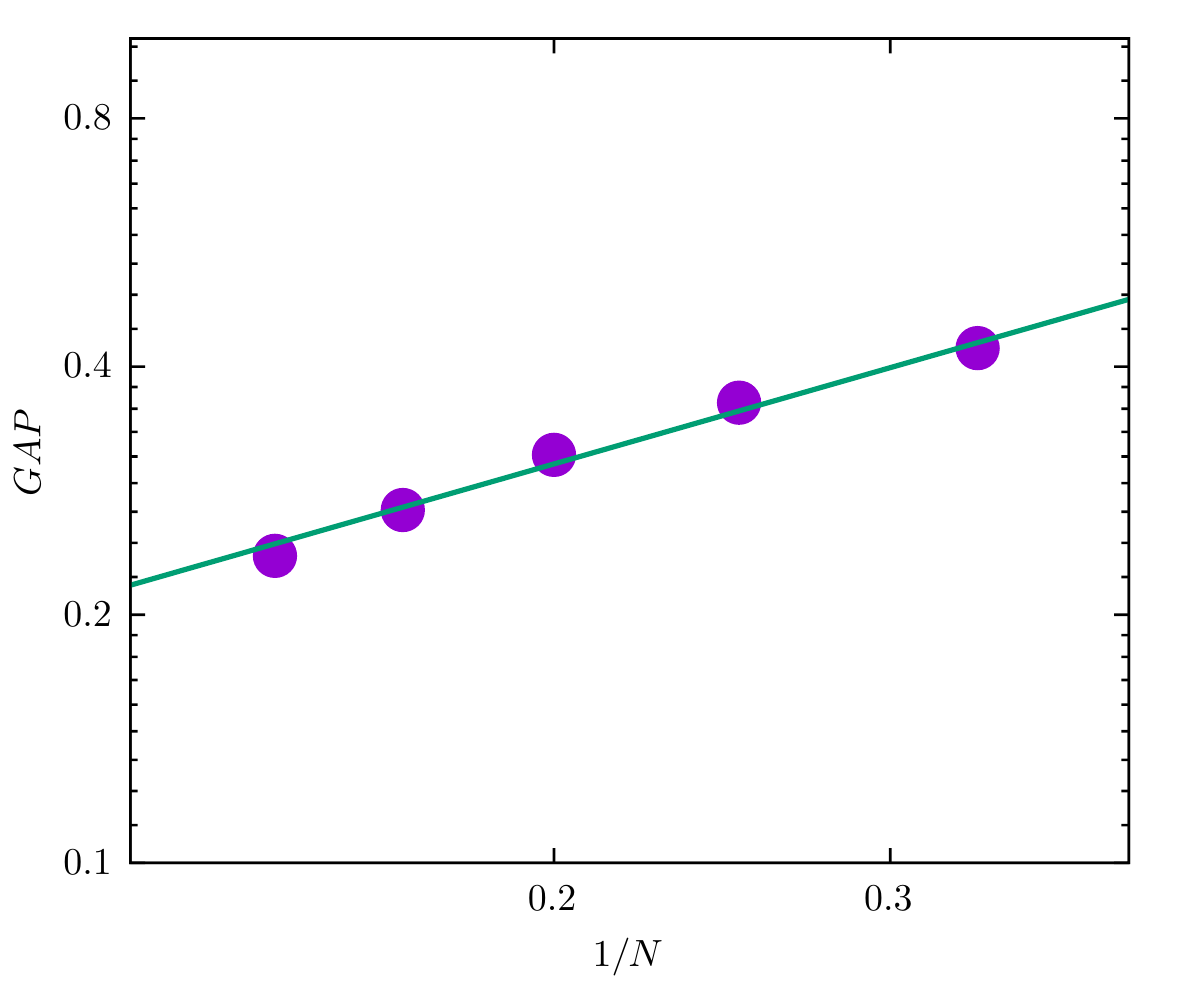}
\par\end{centering}
\caption{Left:  Partial spectrum of the Liouvillian for an $N=3$ and $7$ qubits systems. Right: (logarithmic scale) GAP between the steady-state and the first oscillations coherences as a function of the inverse of the number of qubits. The fit corresponds to the function $f(x)= x^{0.75}$, obtain with the data from \ref{fig5-2}. The parameters of the simulation are $\Omega=5,\;T=0.2$, and one unbiased stored pattern ($P=1$).}
\label{GAP}
\end{figure}

In order to investigate, in a more quantitative way, the damping of oscillations in finite systems and their persistence in infinite systems, we have conducted a spectral analysis of the system's Liouvillian at various sizes. As previously mentioned, oscillations can only endure indefinitely if the Liouvillian's eigenvalues contain pairs with a zero real part and a non-zero imaginary part. In  \ref{GAP} (left), it is shown the spectrum of a Liouvillian associated with QHNNs containing one stored pattern and sizes $N=3,\,7,$ $\Omega=5,\;T=0.2$, and one stored pattern. We observe that the first non-zero eigenvalues form a conjugated pair, and the gap between these pairs and the steady-state narrows as the system size expands.
The right figure displays the gap between these pairs and the steady-state as a function of $1/N$. The gap's trend can be approximated by a power law. The best fit corresponds to the form $\Delta(1/N)=0.886 \,(1/N)^{0.663}$, suggesting that the gap should diminish to zero as $N\to\infty$. This result is also consistent with the fit of the exponential decay of the overlap oscillations illustrated in  \ref{fig5-2}. In fact we can also make a fit of the gap as a function of $x=1/N$ with an exponent of $0.75$ in the form $\Delta(x)\approx x^{0.75}$, fitting which  presents also a very low mean-square error, as it is shown in \ref{GAP}.
Alternative fits, such as a linear function in the form $a+b\,\log(1/N)$, are also acceptable. Consequently, we cannot definitively assert that the gap will close at the thermodynamic limit. However, this is a plausible conclusion based on our simulations, which support the mean field approach of Ref. \cite{rotondo:jpa18}. Unfortunately, observing this effect in quantum computers or simulators will be challenging due to its emergence only in infinite systems.

\begin{figure}[h!]
\begin{centering}
\includegraphics[width=10cm]{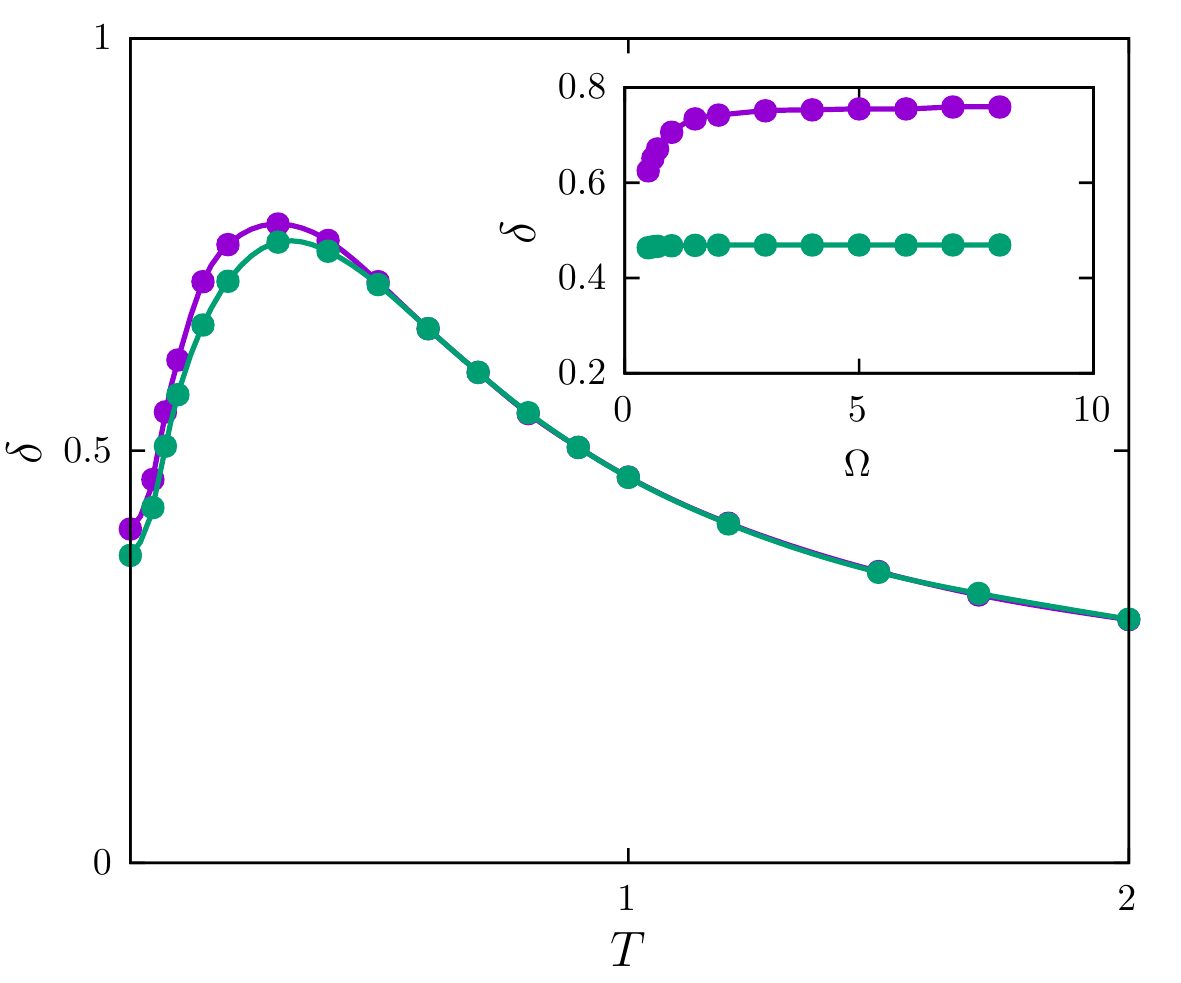}
\par\end{centering}
\caption{Dependence of the scaling parameter $\delta$ with system relevant parameters $T$ and $\Omega.$ (Main figure) Dependence of $\delta$ as a function of $T$ for $\Omega=5$ (purple) and $\Omega=1$ (green). (Inset) Dependence of $\delta$ as a function of $\Omega$ for $T=0.2$ (purple) and $T=1$ (green). Other parameters are $N=8$ and $P=1$ (a single unbiased pattern).}
\label{omegatemp}
\end{figure}

On the other hand, we have also study the dependence of the parameter $\delta$ with relevant parameters of the system, namely $T$ and $\Omega$ as it is illustrated in \ref{omegatemp}. This clearly depicts that the $\delta$ scaling parameter does not depend dramatically on $\Omega$ and $T$ -- in this case the value is finite and confined in the range (0.35,0.75) -- which demonstrates the generality of our findings above.

\subsection{Storage capacity of the quantum Hopfield network}
\begin{figure}[h!]
\begin{centering}
\includegraphics[width=10cm]{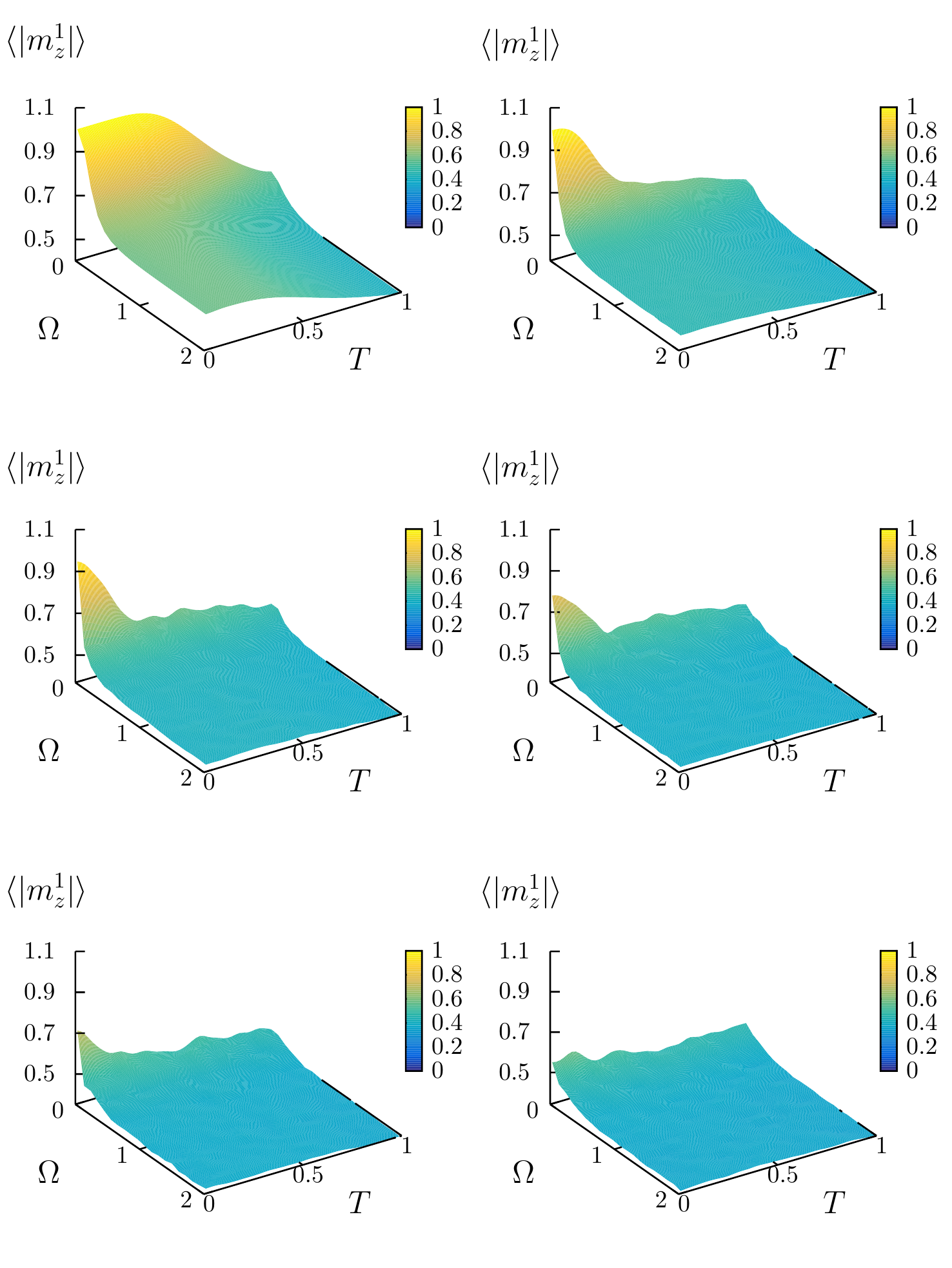}
\par\end{centering}
\caption{Steady state overlap of a system with $N=8$ neurons with the initial state in the ($\Omega,T$) space.  From top to bottom and
from the left to the right overlap surfaces correspond to $P=1,2,3,4,5$
and $6,$ respectively. Any point in the map has been obtained after averaging over 50 pattern realisations.}
\label{fig7}
\end{figure}

A crucial aspect of Hopfield networks is their storage capacity, which refers to the maximum number of distinct patterns or memories that can be stored and accurately recalled. The capacity of a Hopfield network is often represented as a fraction between this number and the number of neurons in the network, $\alpha=P/N$ \cite{gardner:88}. To analyse the capacity of the Quantum Hopfield Neural Network (QHNN), we have investigated the system's behavior as a function of $\Omega$ and $T$ when the number of stored patterns increases, as illustrated in  \ref{fig7}. The simulations are conducted as follows: for each panel of the figure we consider an increasing number of stored patterns (from top to bottom) and set the system to an initial pattern and let it to evolve (for each value of  $\Omega$ and $T$) until it reaches its steady state. We repeat this for $50$ realisations of the stored patterns and average the final state over this number of realisations. For each realisation of stored patterns, we allow the system to evolve using  \ref{lindblad} up to the steady-state, observing the final $Z-$overlap of the system with the initial pattern. Since the system can reach either the pattern or the antipattern, which has the same $Z-$overlap but with an opposite sign, the final state of \ref{lindblad} for the $Z-$overlap operator will be zero. To circumvent this effect, we depict the behavior of the mean value of the operator representing the absolute value of the $Z-$overlap with the initial pattern, $\la |m_{z}^{1}| \ra$.

We aim to highlight that the capacity of the classical Hopfield model is approximately $\alpha \approx 0.138$. This implies that for the network sizes we considered ($N \leq 10$), the system exhibits a ferromagnetic-like phase with only one stored pattern ($P = 1$). With a larger number of patterns, the system transitions into the spin-glass phase. Thus, our findings are particularly pertinent to small-sized networks. Analysing these results to larger, finite-sized networks is highly non-trivial due to the exponential increase in computational cost with larger $N$ values. Additionally, existing mean-field theories for $N \rightarrow \infty$ are limited and fail to adequately reflect the numerical simulation results presented in our study.

In \ref{fig7}, we observe that the QHNN's ability to retrieve the initial pattern declines rapidly as temperature and the quantum parameter $\Omega$ increase. This temperature dependence is analogous to that of the CHNN, where higher temperatures lead to greater thermal fluctuations, thereby reducing the overlap with the recovered pattern. The dependence on $\Omega$ for finite systems is a novel finding, albeit anticipated. As the quantum Hamiltonian causes oscillations within the system, it assists in transitioning between various patterns and anti-patterns. However, this intriguing effect also has the drawback of hampering pattern recovery, even when only a single pattern is present. The issue is exacerbated when multiple patterns are encoded within the system's weights. For instance, when only two patterns are involved, the overlap between the steady state and the initial patterns diminishes, even for relatively low values of $\Omega$. These results are consistent with recent research on infinite systems under a mean-field approach \cite{bodeker:prr23}.

\begin{figure}[h!]
\begin{centering}
\includegraphics[width=12cm]{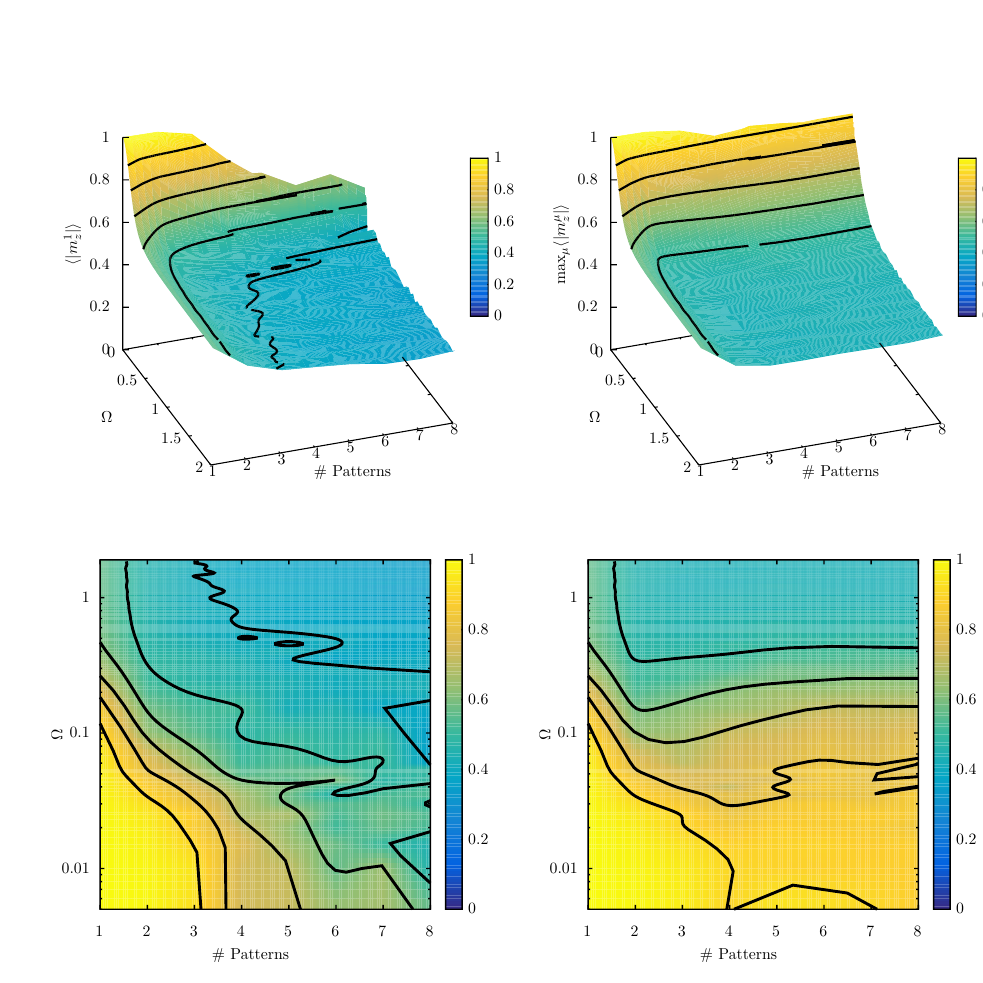}
\par\end{centering}
\caption{Storage capacity of the QHNN as a function of the quantum parameter
$\Omega$ and number of stored patterns. The figure illustrates that when quantum effects are present there is a non-linear effect over the ability of remember a stored pattern as a function of the number of stored patterns. Left panels
corresponds to ability to retrieve the starting pattern (namely $\mu=1$)
whereas the right panels corresponds to ability of the network to
remember or recall any of the other stored patterns. Any point in the map has been obtained after averaging over 50 pattern realisations. All simulations has been computed at very low temperature ($T=0.005$) and with a QHNN of 8 qubits.  The values between the integer number of patterns have been interpolated by a spline method.} 
\label{fig8}
\end{figure}

The examination of storage capacity as temperature approaches zero (i.e., $T\rightarrow0$) is particularly interesting, as it helps mitigate the negative impact of thermal fluctuations on the retrieval of information stored in the attractors. Achieving exactly $T=0$ in simulations is challenging due to the difficulty in defining valid quantum jump operators. However, one can use the lowest possible temperature in simulations. In  \ref{fig8}, we calculate the average steady-state absolute value of the overlap with  stored patterns for a QHNN consisting of 8 qubits at a temperature of $T=0.005$ using the same simulation procedure than in \ref{fig7} but now using as relevant parameters $\Omega$ and $P$. The left panels in  \ref{fig8} illustrate the steady-state absolute $Z-$overlap of the system with the initial pattern as in \ref{fig7}. In contrast, the right panels display the steady-state absolute value of the maximum $Z-$overlap with any of the $P$ stored patterns. Furthermore, and as in \ref{fig7}, each point on the surface in  \ref{fig8} is obtained by averaging over 50 realizations or configurations of stored patterns. This approach avoids any bias due to specific configurations of random patterns used. Note that even for so large number of realizations considered in our work, still is not sufficient to have a very smooth surface in some region of the parameter space. This is a clear indication that we are in a metastable region where sometimes (depending on the realisation) the system recalls a pattern and other times it falls in a mixture state or spin-glass state as in the standard Hopfield model. This fact induce the uncertainly observed in the contour lines of \ref{fig8}.  

 A deep analysis of \ref{fig8} reveals some intriguing phenomena. First for the classical limit ($\Omega=0$) the ability to retrieve the initial pattern is quickly lost when the number of stored patterns $P$ increases (due to the interference among these patterns) and the ability to recall other stored patterns decreases slowly monotonically with large $P$ and which origin can be due to the small size of the system considered and the fact that many of these patterns are very overlapping among them (this effect can be seen by carefully inspecting the behavior of contour lines in bottom panels of \ref{fig8} when approaching to $\Omega=0$). It is worth noting to say that we performed simulations such that the initial pattern has the same number of 1 and 0, but the rest of the stored patterns are unbiased random patterns so the number of 1s and 0s can be different in each pattern realization which introduces variability.  Second, when quantum effects are included $\Omega>0$ and due to the damping effect induced by this term, the ability to retrieve the initial pattern also decreases quickly with $\Omega,$ even for very small $P$. However, there is range of the quantum parameter, i.e. $\Omega\in(0.01,1)$ where although the overlap with the initial pattern decreases quickly with $P$ the capacity of the system to efficiently recall other stored patterns first decreases and after enlarges with increased values of the stored patterns in a non-monotonic way (see this non-monotonic behavior in the contour lines depicted in figure \ref{fig8} bottom right). Consequently, for values of $\Omega$ in this range, there exists a particular number of stored patterns where the  capacity to remember memories reaches a local minimum. Moreover the enhancement of this recall ability of the system for large $P$ occurs even when the number of stored patterns is comparable to the system size.  This is an interesting new feature of the QHNN and may be further investigated for potential applications in machine learning tasks. Finally, for larger values $\Omega>1$, both the retrieval capacity of the initial memory and the recall capacity of any memories of the network diminishes as in the classical limit.

It is also worth noting to say that recent works have treat the storage capacity problem in the present dissipative QHNN under a theoretical point of view (see \cite{Boedeker2023}). In this work the authors assume similar mean-field techniques to those used in \cite{rotondo:jpa18} and additional approaches to analytically compute the storage capacity of the system, for the loading parameter $\alpha>0.$ They reported a similar trend of $\alpha$ in the ($\Omega,T$) parameter space to our results here. However, in absence of large-size simulations it is difficult to validate such mean-field capacity results, which are only valid for $N\rightarrow\infty.$ In fact, our simulations here are the first ones reported in the literature concerning this very interesting issue and only can be done for relatively small system sizes, so one can not conclude anything about these mean-field capacity results.

\section{Conclusions}

In this paper we studied by means extensive numerical simulations
the emergent behavior of a recent reported quantum Hopfield neural
network which has been previously investigated within the framework of
a mean field theory \cite{rotondo:jpa18}. Our motivation was to clarify
if quantum simulations agree with the reported mean-field results in 
finite systems. Additionally, we extended the analysis to study the storage capacity of such quantum Hopfield neural network that until our knowledge has
not been yet reported. All our reported results has been obtained
using the standard Qutip python module. Our main conclusions are the 
following:

\begin{itemize}
\item The oscillations among stored attractors reported in a previous mean-field theory for the long-time regime emerge only in single trajectory simulations. 
\item For finite-size systems, oscillations
among attractors damp in simulations when one averages over many
quantum trajectories. 
\item A study of size-dependent behaviour of the rate of damping of the oscillations and a spectral analysis of the eigenvalues of the Liouvillian indicates that steady state oscillations could occur for infinite size systems, but for any finite size only stochastic changes among metastable attractors in single quantum trajectories and not real limit cycle behaviour emerges.
\item  The inclusion of the quantum term in the system decreases also the probability to be trapped in  mixture or spin-glass states for a long time, particularly for $P$ large, when the quantum effect increases, which eventually makes the system to be more ``ergodic'' like compared with the CHNN.
\item We have also computed the storage capacity of the QHNN and see that
the presence of the quantum term induces instabilities in attractors and therefore decreases the storage capacity to retrieve the initial pattern, but has the ability to retrieve the information encoded in other patterns stored in the system.  The origin of such intriguing non-linear behaviour could be produced by the complex interplay between the inherent initial transient destabilization of the memory attractors induced by the quantum term, the existence of mixture or spin-glass states for large P, and the effect of the damping of the oscillations for larger times that decreases the destabilisation effect of the quantum term in the steady state. This can induce that the system primarily can go out of the initial memory attractor and then can be trapped in other attractor not correlated with the initial one at large times.

\end{itemize}

 It is worth noting to say that our results here are obtained for the case of small size (up to $N=12$). It remains then an open question the effect of the quantum term for larger systems, in which a robust pattern retrieval phase with multiple stored patterns can emerge. Despite the small size considered, our work shows new intriguing phenomena affecting both the retrieval of memories in a pure ferromagnetic phase ($P=1$) and in and non-memory spin-glass phase ($P>1$) which could have strong computational implications if the model is used for machine learning applications. Moreover, our work here is the first rigorous numerical study of a dissipative quantum Hopfield Network and opens new perspectives for future research on quantum neural networks. In particular, it would be of great interest to investigate whether quantum effects improve pattern recognition and classification tasks when the system is in the spin-glass phase. Also the present study can be extended to find quantum encoding mechanisms,  similar to the recently reported in \cite{marsh:prx21}, which could be useful to understand the non-mononotonic behaviour during recall process observed in our study. Finally the damping of pattern-antipattern oscillations observed for finite-size systems and the scaling of the damping time-constant with the system size, in such a way the for infinite systems a coherent oscillatory phase can emerge, resembles the behaviour observed in quantum systems with ``time-crystals'' behaviour \cite{cabot:arxiv22}, which can make the present dissipative QHNN as a very suitable candidate to explore in future works the concept of ``informational time-crystals'' in quantum neural networks.

\section{Acknowledgments}

This paper has been funded by  project PID2021-128970OA-I00 funded by MCIN/AEI/
10.13039/501100011033 and, by ``ERDF A way of making Europe'', by the ``European Union'', the Ministry of
Economic Afairs and Digital Transformation of the Spanish Government through the QUANTUM ENIA
project call - Quantum Spain project, and by the European Union through the Recovery, Transformation and
Resilience Plan - NextGenerationEU within the framework of the Digital Spain 2026 Agenda and FEDER/Junta
de Andaluc\'ia program A.FQM.752.UGR20. J.J.T. acknowledges financial support from the Consejer\'ia de Transformaci\'on Econ\'omica, Industria, Conocimiento y Universidades, Spain, Junta de Andaluc\'ia, Spain and European Regional Development Funds, Ref. $P20_00173$. This work is also part of the Project of I+D+i, Spain Ref. PID2020-113681GBI00, financed by MICIN/AEI/10.13039/501100011033, Spain.

Finally, we are also grateful for the computing resources and related technical support provided by PROTEUS, the supercomputing center of the Institute Carlos I for Theoretical and Computational Physics in Granada, Spain. 

\bibliographystyle{unsrt}


\bibliography{phys.bib}

\end{document}